\documentclass[
reprint,
superscriptaddress,
amsmath,amssymb,
aps,
prc,
]{revtex4-2}

\usepackage{graphicx}
\usepackage{dcolumn}
\usepackage{bm}
\usepackage{multirow}
\usepackage{color} 
\usepackage[breaklinks,colorlinks,linkcolor=blue,citecolor=red,urlcolor=blue]{hyperref} 
\usepackage{color}

\begin{document}

\title{Revisiting the extraction of charge radii of $^{40}$Ca and $^{208}$Pb with muonic atom spectroscopy}

\author{Hui Hui Xie}
\affiliation{College of Physics, Jilin University, Changchun 130012, China}

\author{Tomoya Naito}
\affiliation{RIKEN Interdisciplinary Theoretical and Mathematical Sciences Program, Wako 351-0198, Japan}
\affiliation{Department of Physics, Graduate School of Science,	The University of Tokyo, Tokyo 113-0033, Japan}

\author{Jian Li}\email{jianli@jlu.edu.cn}
\affiliation{College of Physics, Jilin University, Changchun 130012, China}

\author{Haozhao Liang}\email{haozhao.liang@phys.s.u-tokyo.ac.jp}
\affiliation{Department of Physics, Graduate School of Science,	The University of Tokyo, Tokyo 113-0033, Japan}
\affiliation{RIKEN Interdisciplinary Theoretical and Mathematical Sciences Program, Wako 351-0198, Japan}

\date{\today}

\begin{abstract}
The extractions of nuclear charge radii from muonic atom spectroscopy for $^{40}$Ca and $^{208}$Pb are revisited to analyze the model dependencies induced by employing a Fermi-type charge distribution.
For that, the charge densities, together with the corresponding muonic transition energies, calculated by the covariant density functional theory are used as a benchmark.
The root-mean-square deviation of transition energies is calculated to quantitatively investigate the sensitivities of transition energies to the details of the two-parameter Fermi distribution.
It is found that the second and fourth moments of the charge distribution can be extracted accurately from the muonic atom spectroscopy without much model dependencies, whereas the obtained two-parameter Fermi distributions cannot reproduce the details of the benchmarking charge densities and, in particular, its surface-diffuseness parameter cannot be determined accurately with the present experimental uncertainties on the muonic transition energies.
\end{abstract}

\maketitle


\section{Introduction}

Muonic atoms, where a negatively charged muon is captured by a nucleus, have been widely used to extract  nuclear charge radii with high-precision for most stable isotopes~\cite{ENGFER1974509,PhysRevC.23.533,PhysRevC.37.2821,SCHALLER1978225,PhysRevC.18.1474,PhysRevC.39.195}, since the technique of X-ray spectroscopy from $\mu$-mesonic atoms developed in 1953~\cite{PhysRev.92.789}. In addition, nuclear electric quadrupole moments can also be accurately deduced from the observed hyperfine splittings of muonic X-ray transitions~\cite{PhysRevC.1.1184,steffen1985precision,PhysRevC.101.054313}. However, there exists a long-standing problem that
limits the extractive accuracy and leads to the non-negligible theoretical uncertainty, when the experimental transition energies are analyzed, i.e., there exists a puzzling discrepancy between theoretical calculations and experimental data for fine-structure splittings~\cite{PhysRevLett.42.1470,PhysRevC.37.2821}.

In the early analysis, the nuclear polarization (NP) correction is supposed to be the main source of such a discrepancy, due to the existence of challenging problems in  the calculation of the NP effect~\cite{PhysRevLett.42.1470,PhysRevC.37.2821}. Thus, considerable theoretical efforts have been made in recent decades to perform a rigorous calculation of NP correction ~\cite{TANAKA1994291,PhysRevC.75.044315,HAGA200571,HAGA2005234,PhysRevLett.128.203001}. Employing state-of-the-art techniques, the most comprehensive calculations of NP corrections are performed recently by Valuev~\textit{et al.}~\cite{PhysRevLett.128.203001} and it is concluded that the NP effect is unlikely responsible for the discrepancy. As a result, the other effects could play roles in explaining the discrepancy.
On the other hand, note that the influence of model dependencies, induced by employing a Fermi-type charge distribution in the model-dependent analysis, is in general not examined~\cite{PhysRevC.37.2821}. Although the influence of model dependencies could be evaded in a model-independent combined analysis of muonic data and electron-scattering data as shown in Ref.~\cite{PhysRevLett.42.1470}, it is also important to investigate the contribution of model dependencies to the uncertainty of  extracted charge radii. For this purpose, a detailed analysis of model dependencies with respect to the Fermi-type charge distribution is performed in the present work by using the theoretical charge density distribution from the self-consistent mean-filed calculations as a benchmark. 

In recent decades, the covariant (relativistic) density functional theory (CDFT) has attracted extensive attentions on account of its successful description of many nuclear phenomena~\cite{RING1996193,MENG2006470,Meng_2015,VRETENAR2005101,NIKSIC2011519,meng2013progress}. For instance, it can well reproduce the isotopic shifts in the Pb~\cite{SHARMA19939} and Sn~\cite{PhysRevC.107.054307} regions, and naturally give the origins of the pseudospin symmetry in the nucleon spectrum~\cite{LIANG20151} and the spin symmetry in the anti-nucleon spectrum~\cite{PhysRevLett.91.262501,liang2010spin}. 
In addition, it can provide a consistent description of currents and time-odd fields, which play an important role in nuclear magnetic moments~\cite{Li_2020,Li2011,10.1143/PTP.125.1185,10.1143/PTPS.196.400,PhysRevC.88.064307} and nuclear rotations~\cite{PhysRevLett.71.3079,PhysRevLett.107.122501}.
Based on the CDFT with the Bogoliubov transformation in the coordinate representation, the relativistic continuum Hartree-Bogoliubov (RCHB) theory was developed~\cite{PhysRevLett.77.3963,Meng1998NPA} to provide a proper treatment of pairing corrections and mean-filed potentials in the presence of the continuum and has achieved great success in various aspects about exotic nuclei, such as providing a microscopic and self-consistent description of halo in $^{11}$Li~\cite{PhysRevLett.77.3963}, predicting the giant halos phenomena in light and medium-heavy nuclei~\cite{PhysRevLett.80.460,PhysRevC.65.041302,Shuang_Quan_2002}, and reproducing the interaction cross sections and charge-changing cross sections in sodium isotopes and other light exotic nuclei~\cite{MENG19981,MENG2002209}.

In this work, nuclear charge densities from the RCHB calculations are used as a benchmark to revisit the extractive process of charge radii based on two-parameter Fermi (2pf) distribution for two doubly-magic nuclei, $^{40}$Ca and $^{208}$Pb. In addition to the second moment, the fourth moment of nuclear charge distribution has recently attracted great attention for various applications, such as providing information on nuclear surface properties~\cite{PhysRevC.101.021301} and performing a successful study of the neutron-skin thickness~\cite{PhysRevC.104.024316}. In this paper, the sensitivities of muonic transition energies to the second and fourth moments are also investigated.

The paper is organized as follows. In Sec.~\ref{sec-2}, we introduce the RCHB method for solving the nuclear structure and then construct the nuclear charge densities. In Sec.~\ref{sec-3-a}, we determine the parameters of 2pf distribution by fitting the charge densities from RCHB calculations in several ways and compare the corresponding transition energies with the data of RCHB. In Sec.~\ref{sec-3-b}, we determine the parameters by fitting the referenced transition energies, namely the results with the charge densities from RCHB calculations. In this way, we qualitatively and quantitatively investigate the model dependencies and the sensitivities of transition energies to the second and fourth moments. Then, an analysis of the Barrett model is also made for further comparison. Finally, summary and perspectives are presented in Sec.~\ref{sec-4}.

\section{Theoretical framework}\label{sec-2}

\subsection{Nuclear charge density from RCHB theory}

In this study, the RCHB theory constructed with the contact interaction in the point-coupling representation between nucleons is adopted, where the conventional finite-range meson-exchange interaction is replaced by the corresponding local four-point interaction between nucleons. The details of the RCHB theory with point-coupling density functional can be found in Refs.~\cite{PhysRevC.82.054319,Xia2018}. In the following, we introduce the theoretical framework briefly.

Starting from the Lagrangian density, the energy density functional of the nuclear system can be constructed under the mean-field and no-sea approximations.
By minimizing the energy density functional with respect to the
densities, one obtains the Dirac equation for nucleons within the
relativistic mean-field framework~\cite{Meng2015}. The relativistic Hartree-Bogoliubov model provides a unified description of both the mean field and the pairing correlation.

In the RCHB theory, the proton and neutron densities can be constructed by quasiparticle wave functions,
\begin{equation}\label{eq:rhov}
	\rho_{\tau}(\bm r)   =\sum_{k\in\tau } V_k^{\dagger}(\bm r)V_k(\bm r)
\end{equation}
with $\tau\in\{p,n\}$.
The nuclear charge density then includes the contributions from the point neutron density, the proton and neutron spin-orbit densities, and the single-proton and single-neutron charge densities, in addition to that from the point proton
density~\cite{Friar1975,PhysRevC.62.054303,PhysRevC.103.054310,kurasawa2019n}. The relativistic nuclear charge density is finally written as
\begin{equation}\label{eq-rhoc}
	\rho_c(r)=\sum_{\tau}\left[\rho_{c\tau}(r)+W_{c\tau}(r)\right],
\end{equation}
where
\begin{align}
	&\rho_{c\tau}(r)=\frac1r\int_0^\infty x\rho_\tau(x)\left[g_\tau(|r-x|)-g_\tau(r+x)\right]\mathrm dx,\\
	&W_{c\tau}(r)=\frac1r\int_0^\infty xW_\tau(x)\left[f_{2\tau}(|r-x|)-f_{2\tau}(r+x)\right]\mathrm dx.
\end{align}
Here, $\rho_\tau(r)$ is the point nucleon density in Eq.~(\ref{eq:rhov}) and $W_\tau(r)$ is the spin–orbit density given in Refs.~\cite{PhysRevC.62.054303,kurasawa2019n}. The functions $g_\tau(x)$ and $f_{2\tau}(x)$ are given by
\begin{eqnarray}
	g_\tau(x)=\frac1{2\pi}\int_{-\infty}^{\infty}e^{iqx}\, G_{E\tau}(\bm q^2)\,\mathrm dq,\\
	f_{2\tau}(x)=\frac{1}{2\pi}\int_{-\infty}^{\infty}e^{iqx}\, F_{2\tau}(\bm q^2)\,\mathrm dq,
\end{eqnarray}
in which $G_{E\tau}$ and $F_{2\tau}$ denote the electric Sachs and Pauli form factors of a nucleon, respectively. The following forms~\cite{PhysRevC.62.054303} are used in this study,
\begin{align}\label{eq-form-factor}
	G_{Ep}(q^2) &=\frac{1}{\left(1+r_p^2\bm q^2/12\right)^2}, \nonumber\\
    G_{En}(q^2) &=\frac{1}{\left(1+r_+^2\bm q^2/12\right)^2}-\frac{1}{\left(1+r_-^2\bm q^2/12\right)^2},\nonumber\\
	F_{2p}(q^2) &=\frac{G_{Ep}}{1+\bm q^2/4M_p^2},\nonumber\\
   F_{2n}(q^2) &=\frac{G_{Ep}-G_{En}/\mu_n}{1+\bm q^2/4M_n^2},
\end{align}
with the proton charge radius $r_p= 0.8414$ fm~\cite{RevModPhys.93.025010} and $r^2_\pm= r_{\mathrm{av}}^2\pm \frac12 \langle r_n^2\rangle $, where $r_{\mathrm{av}}^2= 0.9$~fm$^2$ is the average of the squared radii for positive and negative charge distributions, while $\langle r_n^2\rangle =-0.11$~fm$^2$~\cite{atac2021measurement} is the mean squared charge radius of a neutron. See reference~\cite{PhysRevA.107.042807} for more details.

\subsection{Dirac equation for muonic atom}

The Dirac equation for a muon reads
\begin{equation}\label{eq-dirac-one-electron}
	\left[\bm\alpha\cdot \bm p+\beta M_r+V(\bm r)\right]\psi(\bm r)=\varepsilon \psi(\bm r),
\end{equation}
where the eigenenergy $\varepsilon$ includes both the muonic atom energy levels $E$ and reduced mass $M_r$, i.e., 
\begin{equation}\label{reduced-mass}
  \varepsilon \simeq E+M_r, \quad \mbox{with} \quad  M_r=\frac{M_Am_\mu}{M_A+m_\mu}.
\end{equation}
Note that Eq.~\eqref{reduced-mass} holds exactly only at the non-relativistic limit, while it is a good enough approximation in the present calculations. The mass of muon $m_\mu=206.7682830\,m_e$ is adopted~\cite{RevModPhys.93.025010} and nuclear mass $M_A$ is obtained from Ref.~\cite{Wang_2021}. Since the electrostatic potential $V(r)$ made by the atomic nucleus is spherically symmetric, the eigenvalue function can be written as
\begin{equation}\label{eq-dirac-s}
	\psi_{n\kappa m}(\bm r)=\frac1r\begin{pmatrix}
		iP_{n\kappa}(r)Y_{jm}^l(\theta,\varphi)\\
		Q_{n\kappa}(r)\left(\bm\sigma\cdot\bm{\hat r}\right)Y_{jm}^l(\theta,\varphi)
	\end{pmatrix},
\end{equation}
where $P_{n\kappa}(r)$ and $Q_{n\kappa}(r)$ are its large and small components, respectively.

The radial Dirac equation for one-muon system can be then written in the form of
\begin{align}\label{eq-dirac}
	\begin{pmatrix}
		V(r) & -\left(\frac{\mathrm d}{\mathrm dr}-\frac{\kappa}{ r}\right)\\
		\left(\frac{\mathrm d}{\mathrm dr}+\frac \kappa r\right) &  V(r)-2M_r
	\end{pmatrix}
	\begin{pmatrix}
		P_{n\kappa}(r)\\Q_{n\kappa}(r)
	\end{pmatrix}&=
	E\begin{pmatrix}
		P_{n\kappa}(r)\\Q_{n\kappa}(r)
	\end{pmatrix},
\end{align}
where the mass term $M_r$ is subtracted on both sides of the equation.  

The electrostatic potential $V(r)$ is obtained via~\cite{engel2002relativistic}
\begin{equation}\label{eq-v}
	V(r)=-4\pi\alpha\left[\int_0^r\rho_c(r')\frac{r'^2}{r}\,\mathrm dr'+\int_r^\infty\rho_c(r')r'\,\mathrm dr'\right],
\end{equation}
where $\alpha$ is the fine structure constant.
In the present work, the muonic energy levels are evaluated numerically.

\section{Results and discussion}\label{sec-3}

In order to investigate the influence of model dependency on muonic atom spectrum, nuclear charge densities of $^{40}$Ca and $^{208}$Pb obtained from the RCHB calculations are used as benchmarks. The relativistic density functional PC-PK1~\cite{PhysRevC.82.054319}, which provides one of the best density-functional descriptions for nuclear properties~\cite{RN7,PhysRevC.91.027304,PhysRevC.86.064324,YAO2013459,LI2013866,LI2012470,PhysRevC.88.057301,Wang_2015}, is employed. The box size $R_{\rm box}=20$~fm, the mesh size $\Delta r=0.1$~fm, and the angular momentum cutoff $J_{\rm max}=19/2~\hbar$ are used in the RCHB calculations. More numerical details can be found in Ref.~\cite{Xia2018}.

The Dirac equation for the muon is solved using the generalized pseudospectral (GPS) method, whose powerful performance has been shown in Refs.~\cite{canuto2007spectral,PhysRevA.104.022801,https://doi.org/10.1002/qua.26653,Jiao_2021}. The mapping parameters $L=0.01$ and $k=3$ are used in the GPS calculations.

\subsection{Fitting the nuclear charge density}\label{sec-3-a}

In general, the available experimental data of muonic atom spectrum are analyzed by means of a 2pf distribution in the form of
\begin{equation}
	\rho_c(r)=\frac{\rho_0}{1+\exp\left[4\ln3(r-c)/t\right]},
\end{equation}
with the half-density-radius $c$ and surface-diffuseness parameters $t$, together with the normalization
\begin{equation}
	4\pi\int_0^\infty \rho_c(r) r^2\, \mathrm dr=Z,
\end{equation}
where $Z$ represents the proton number. 
The $n$-th moment is given by
\begin{equation}
	R_n\equiv\langle r^n\rangle=\frac{4\pi}{Z}\int_0^\infty\rho_c(r) r^{n+2}\, \mathrm dr.
\end{equation}

In the following investigation, first of all, the parameters $c$ and $t$ of 2pf distribution are determined by fitting the RCHB charge densities in several feasible ways, \textcolor{black}{with $\rho_0$ determined by the normalization.}
\textcolor{black}{Specifically, four 2pf distributions are obtained here, i.e., 2pf($\sigma_1$) is determined by minimizing the loss function $\sigma_1$, 2pf($R_2$,$\sigma_1$) and 2pf($R_2$,$\sigma_2$) are determined by minimizing respectively $\sigma_1$ and $\sigma_2$ with the second moment $R_2$ constraint, and 2pf($R_2$,$R_4$) is determined by constraining the second and fourth moments.}
The loss functions $\sigma_1$ and $\sigma_2$ are two alternatives that indicate the deviations in shape with the following forms:
\begin{equation}
	\sigma_1=\int_0^{\infty}\left(\rho_{c}^\mathrm{2pf}(r)-\rho_{c}^\mathrm{RCHB}(r)\right)^2\, \mathrm dr
\end{equation}
and
\begin{equation}
	\sigma_2=\int_0^{\infty}\left(\rho_{c}^\mathrm{2pf}(r)-\rho_{c}^\mathrm{RCHB}(r)\right)^2 r^2\, \mathrm d r.
\end{equation}

\begin{figure*}
	\centering
	\includegraphics[width=17cm]{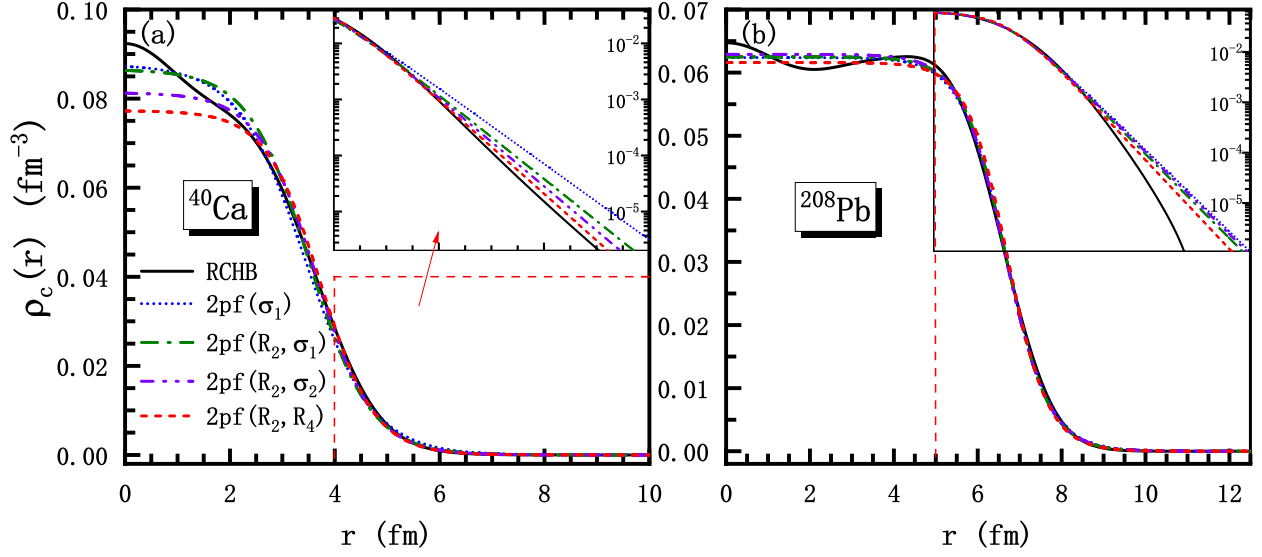}
	\caption{Comparison of charge densities from the RCHB theory with those of 2pf distributions for (a) $^{40}$Ca and (b) $^{208}$Pb. The 2pf parameters are determined by fitting the RCHB charge densities in four different ways, i.e., 2pf($\sigma_1$), 2pf($R_2$,$\sigma_1$), 2pf($R_2$,$\sigma_2$), and 2pf($R_2$,$R_4$). See texts for details.}
	\label{fig:fig1}
\end{figure*}

As shown in Fig.~\ref{fig:fig1}, four 2pf distributions are  compared with the charge density from the RCHB calculations.
It can be seen from Fig.~\ref{fig:fig1}(a) that the central densities of four 2pf distributions for $^{40}$Ca are visibly different. The curve of 2pf($R_2$,$R_4$) deviates furthest from the RCHB calculation, and the curve of 2pf($R_2,\sigma_2$) is the next. In contrast, the curve of 2pf($\sigma_1$) is closest in shape to the RCHB calculation. As for $^{208}$Pb in Fig.~\ref{fig:fig1}(b), it shows less differences between four 2pf distributions. Nevertheless, the curve of 2pf($R_2$,$R_4$) also shows the most deviation from the RCHB calculation. On the other hand, in the insets of Fig.~\ref{fig:fig1} the logarithmic coordinates are taken to emphasize the difference of charge density at the surface. One can see from the insets  that the distributions of 2pf($R_2$,$R_4$) are the closest ones to the RCHB charge density distributions among  four distributions at the nuclear surface, i.e., $r>6$~fm for $^{40}$Ca and $r>8$~fm for $^{208}$Pb. In contrast, the charge distributions 2pf($\sigma_1$) are in the worst agreement with the RCHB calculations. With regard to the rest of two 2pf distributions, the curve of 2pf($R_2$,$\sigma_2$) is closer to (further away from) the curve of RCHB than 2pf($R_2$,$\sigma_1$) for $^{40}$Ca ($^{208}$Pb).

The fitted parameters $c$ and $t$ of four 2pf distributions, together with the corresponding second and fourth moments and charge radii $r_c$, \textcolor{black}{as well as two loss functions $\sigma_1$ and $\sigma_2$} are shown in Table~\ref{tab-1}. Since the results of 2pf($\sigma_1$) are obtained without the second-moment constraint, there exists a visible difference in the second moment between 2pf($\sigma_1$) and the others, i.e., around $0.7$~fm$^2$ for $^{40}$Ca and $0.2$~fm$^2$ for $^{208}$Pb. Moreover, it can be seen that the parameters determined in 2pf($\sigma_1$), 2pf($R_2$,$\sigma_1$), and 2pf($R_2$,$\sigma_2$), without the fourth-moment constraint, generally overestimate the fourth moments.

\begin{table*}[!ht]
	\caption{The parameters $c$ and $t$, the second and fourth moments $R_2$ and $R_4$, and the charge radii $r_c$, \textcolor{black}{as well as the values of two loss functions $\sigma_1$ and $\sigma_2$}, for $^{40}$Ca and $^{208}$Pb in four 2pf models. Note that the numbers that exactly reproduce the RCHB values are highlighted in bold form.}\label{tab-1}
	\centering
		\begin{tabular}{c|cccc|cccc}
			\toprule
			&                           \multicolumn{4}{c}{$^{40}$Ca}\vline                            &                           \multicolumn{4}{c}{$^{208}$Pb}                           \\ \hline
			& 2pf($\sigma_1$) & 2pf($R_2$,$\sigma_1$) & 2pf($R_2$,$\sigma_2$) & 2pf($R_2$,$R_4$) & 2pf($\sigma_1$) & 2pf($R_2$,$\sigma_1$) & 2pf($R_2$,$\sigma_2$) & 2pf($R_2$,$R_4$) \\ \hline
			$c$ (fm)   & $3.43431$         & $3.51641$               & $3.63210$               & $3.72839$                & $6.65823$         & $6.66207$               & $6.64228$               & $6.70918$          \\
			$t$ (fm)   &$ 2.82022$         & $2.55115$               & $2.41134$               & $2.28476$                & $2.33968$         & $2.26797$               & $2.31614$               & $2.14837$          \\ \hline
			$R_2$ (fm$^2$) & $12.7666$         & $\textbf{12.0755}$               & $\textbf{12.0755}$               & $\textbf{12.0755}$                & 30.5160         & $\bf 30.3103$               & $\bf 30.3103$               & $\bf 30.3103$          \\
			$R_4$ (fm$^4$) & $256.054$         & $220.280$               & $214.499$               & { $\textbf{209.699}$}                & $1195.877$        & $1174.858$              & $1178.577$              & $\bf 1166.040$         \\
			$r_c$ (fm) & $3.5730$          &  $\textbf{3.4750}$                & $\textbf{3.4750}$                & $\textbf{3.4750}$                 & $5.5241$          & $\bf 5.5055$                & $\bf 5.5055$                & $\bf 5.5055$           \\ 
   $\sigma_1$ ($10^{-6}$fm$^{-5}$)&$32.1726$&$44.8123$&$90.7370$&$195.203$&$9.71232$&$10.42702$&$11.3099$&$15.5466$\\
   $\sigma_2$ ($10^{-5}$fm$^{-3}$)&$19.1357$&$18.4373$&$9.36914$&$16.1918$&$10.4874$&$13.9824$&$12.6133$&$29.1877$\\
   \toprule
		\end{tabular}
\end{table*}

\begin{table*}[!ht]
	\caption{Muonic transition energies (keV) calculated by using RCHB charge density, and the differences between them and the results of four 2pf distributions, for $^{40}$Ca and $^{208}$Pb. Numbers in parentheses represent the power of ten.}\label{tab-2}
	\centering
		\begin{tabular}{c|r|lrrr|r|lrrr}
			\toprule
			Transition           & \multirow{1}{*}{$E_{\mathrm{RCHB}}$} & \multicolumn{4}{c}{$^{40}$Ca, $\Delta E_{\mathrm{2pf}}-\Delta E_{\mathrm{RCHB}}$}\vline& \multirow{1}{*}{$E_{\mathrm{RCHB}}$} & \multicolumn{4}{c}{$^{208}$Pb, $\Delta E_{\mathrm{2pf}}-\Delta E_{\mathrm{RCHB}}$} \\
			(keV)             &         ($^{40}$Ca)              & 2pf($\sigma_1$) & 2pf($R_2$,$\sigma_1$) & 2pf($R_2$,$\sigma_2$) & 2pf($R_2$,$R_4$) &       ($^{208}$Pb)                & 2pf($\sigma_1$) & 2pf($R_2$,$\sigma_1$) & 2pf($R_2$,$\sigma_2$) & 2pf($R_2$,$R_4$) \\ \hline
			$2p_{3/2}$-$1s_{1/2}$      & $778.229$              & $-2.58$         & $1.82(-1)$          & $8.47(-2)$          & $3.69(-3)$   	       & $5917.293$              & $-8.77$         & $4.84$\quad\;\;\;\;              & $7.06$\quad\;\;\;\;              & $-4.51(-1)$  \\
$2p_{1/2}$-$1s_{1/2}$      & $776.766$              & $-2.58$         & $1.82(-1)$          & $8.47(-2)$          & $3.68(-3)$	       & $5735.444$              & $-8.24$         & $4.67$\quad\;\;\;\;              & $6.82$\quad\;\;\;\;              & $-4.57(-1)$      \\
$3d_{3/2}$-$2p_{1/2}$     & $157.439$              & $-8.03(-3)$     & $-1.06(-3)$           & $-4.46(-4)$           & $7.14(-5)$      & $2619.561$              & $-2.69$         & $-7.98(-4)$           & $-9.02(-2)$           & $2.15(-1)$      \\
$3d_{5/2}$-$2p_{3/2}$     & $156.124$              & $-5.43(-3)$     & $-1.11(-3)$           & $-4.74(-4)$           & $5.95(-5)$      & $2479.793$              & $-2.20$         & $-1.78(-1)$           & $-3.40(-1)$           & $2.12(-1)$    \\
$3d_{3/2}$-$2p_{3/2}$     & $155.976$              & $-5.43(-3)$     & $-1.11(-3)$           & $-4.74(-4)$           & $5.96(-5)$      & $2437.712$              & $-2.16$         & $-1.72(-1)$           & $-3.31(-1)$           & $2.09(-1)$     \\
$5f_{7/2}$-$3d_{5/2}$      & $79.851$                & $-1.96(-6)$     & $-6.14(-7)$           & $-3.05(-7)$           & $-5.66(-8)$     & $1358.683$              & $-5.67(-2)$     & $-1.88(-2)$           & $-2.75(-2)$           & $1.87(-3)$   \\
$5f_{5/2}$-$3d_{5/2}$      & $79.835$                & $-1.95(-6)$     & $-6.05(-7)$           & $-2.96(-7)$           & $-4.78(-8)$     & $1354.023$              & $-5.64(-2)$     & $-1.87(-2)$           & $-2.74(-2)$           & $1.87(-3)$  \\
$4f_{7/2}$-$3d_{5/2}$      & $54.599$                & $-1.96(-6)$     & $-6.07(-7)$           & $-2.99(-7)$           & $-5.04(-8)$     & $932.086$               & $-5.66(-2)$     & $-1.88(-2)$           & $-2.74(-2)$           & $1.87(-3)$    \\
$4f_{5/2}$-$3d_{5/2}$      & $54.568 $               & $-1.95(-6)$     & $-6.06(-7)$           & $-2.98(-7)$           & $-4.95(-8)$     & $923.018$               & $-5.63(-2)$     & $-1.87(-2)$           & $-2.73(-2)$           & $1.87(-3)$    \\
$4d_{5/2}$-$3p_{1/2}$     & $55.074$                & $-2.81(-3)$     & $-3.70(-4)$           & $-1.55(-4)$           & $2.53(-5)$       & $914.494$               & $-7.84(-1)$     & $4.73(-2)$          & $4.00(-2)$          & $6.57(-2)$ \\
$4d_{5/2}$-$3p_{3/2}$     & $54.705$                & $-1.91(-3)$     & $-3.90(-4)$           & $-1.66(-4)$           & $2.13(-5)$       & $885.170$               & $-6.94(-1)$     & $-2.38(-2)$           & $-6.41(-2)$           & $7.31(-2)$  \\
$4d_{3/2}$-$3p_{3/2}$     & $54.643$                & $-1.91(-3)$     & $-3.90(-4)$           & $-1.66(-4)$           & $2.13(-5)$       & $867.687$               & $-6.75(-1)$     & $-2.12(-2)$           & $-5.96(-2)$           & $7.15(-2)$    \\ 		
			\toprule
		\end{tabular}
\end{table*}

In Table~\ref{tab-2}, the muonic transition energies calculated by using the RCHB charge density, and the differences between them and the results calculated by using the 2pf distributions, for $^{40}$Ca and $^{208}$Pb, are shown. Twelve muonic transition energies for $^{40}$Ca and $^{208}$Pb are given, namely, $2p_{3/2}$-$1s_{1/2}$, $2p_{1/2}$-$1s_{1/2}$, $3d_{3/2}$-$2p_{1/2}$, $3d_{5/2}$-$2p_{3/2}$, $3d_{3/2}$-$2p_{3/2}$, $5f_{7/2}$-$3d_{5/2}$, $5f_{5/2}$-$3d_{5/2}$, $4f_{7/2}$-$3d_{5/2}$, $4f_{5/2}$-$3d_{5/2}$, $4d_{5/2}$-$3p_{1/2}$, $4d_{5/2}$-$3p_{3/2}$, and $4d_{3/2}$-$3p_{3/2}$, involving fourteen levels. 

In comparison, the results of 2pf($\sigma_1$) are in the worst agreement with RCHB calculations and the results of  2pf($R_2$,$R_4$) are in the best agreement with RCHB calculations.
For example, the 2pf($\sigma_1$)  gives $2.58$~keV derivation from RCHB for $2p_{3/2}$-$1s_{1/2}$ in $^{40}$Ca. After considering the second moment constraint $R_2$ with different loss function minimized, 2pf($R_2$,$\sigma_1$) and 2pf($R_2$,$\sigma_2$) give the derivations $-0.182$ and $-0.0847$~keV, respectively. Furthermore, with both second- and fourth-moment constraints, the derivation from RCHB calculations decreases to $-3.65\times10^{-3}$~keV. A similar conclusion for other transition energies in $^{40}$Ca and $^{208}$Pb is also observed in Table~\ref{tab-2}.
In view of the comparison of charge distribution between RCHB calculations and four 2pf distributions shown in Fig.~\ref{fig:fig1}, it is concluded that the surface of nuclear charge density plays a much more important role in the muonic atom spectrum than the center. 
Furthermore, it implies the considerable sensitivities of the transition energies to the second and fourth moments. A quantitative discussion is given in the next section.

\subsection{Fitting the muonic transition energies}\label{sec-3-b}

In this section, the procedure to extract the charge radii from the muonic transition energies with the 2pf distribution will be discussed. For the sake of discussion, the root-mean-square deviation (RMSD) for muonic transition energies is defined as
\begin{equation}\label{eq-chi}
	\delta=\left[\frac{1}{N}\sum_i\left(\Delta E_{i}^\mathrm{2pf}-\Delta E_{i}^\mathrm{RCHB}\right)^2\right]^{1/2},
\end{equation}
where $E_i^{\mathrm{2pf}}$ and $E_i^{\mathrm{RCHB}}$ are the $i$-th transition energy calculated with the 2pf charge distributions and the RCHB charge distributions, respectively, and $N$ is the total number of transitions considered. For convenience, four transitions with the largest transition energies, i.e., $2p_{3/2}$-$1s_{1/2}$,  $2p_{1/2}$-$1s_{1/2}$, $3d_{3/2}$-$2p_{1/2}$, and $3d_{5/2}$-$2p_{3/2}$, are considered in Eq.~(\ref{eq-chi}).

\begin{figure*}
	\centering
	\includegraphics[width=17cm]{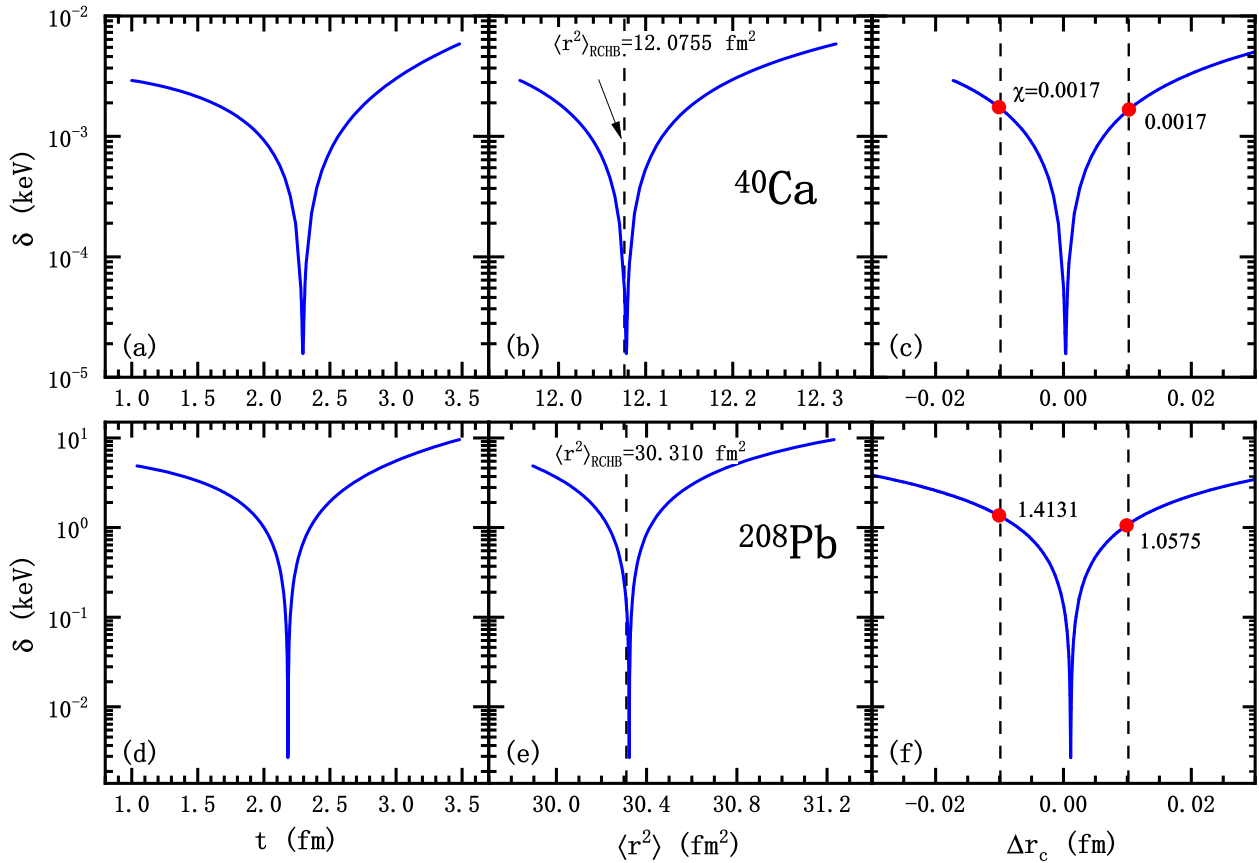}
	\caption{The variation of $\delta$ for $^{40}$Ca and $^{208}$Pb: (a) and (d) $\delta$ versus parameter $t$ of 2pf distribution, where parameter $c$ are determined to minimize $\delta$; (b) and (e) $\delta$ versus the second moment of 2pf distribution, where the black dashed lines indicate the second moment of RCHB charge distributions; (c) and (f) $\delta$ versus the difference of charge radii between the 2pf distributions and RCHB calculations, i.e., $\Delta r_c=r_{c}^\mathrm{2pf}-r_{c}^\mathrm{RCHB}$, where the black dashed lines represent $\Delta r_c=\pm0.01$~fm.}
	\label{fig:fig2}
\end{figure*}

In panels (a) and (d) of Fig.~\ref{fig:fig2}, the trend of $\delta$ with the parameter $t$ changing from $1$ to $3.5$~fm is shown, where the parameter $c$ is determined by minimizing \textcolor{black}{the RMSD} $\delta$ for each $t$. 
\textcolor{black}{A minimum point of $\delta$ can be seen in the panels. Here, we use the shorthand writing 2pf($\Delta E$) to denote the best-fit 2pf distribution which yield a minimum value of $\delta$.} 
The minimum values of $\delta$ read $\delta \simeq 2.0\times10^{-5}$~keV with $t=2.2950$~fm and $c=3.7214$~fm for $^{40}$Ca and $\delta \simeq 0.0031$~keV with $t=2.1821$~fm and $c=6.6977$~fm for $^{208}$Pb. 
In addition, $\delta$ versus the second moment $\langle r^2\rangle$ \textcolor{black}{of the 2pf distribution with the same parameters as Figs.~\ref{fig:fig2}(a) and \ref{fig:fig2}(d)} is also displayed in Figs.~\ref{fig:fig2}(b) and \ref{fig:fig2}(e). 
The second \textcolor{black}{and fourth moments} of the best fit are determined as the point of minimum $\delta$, namely $\langle r^2\rangle=12.0778$~fm$^2$, $\langle r^4\rangle=210.145$~fm$^4$ for $^{40}$Ca, and $\langle r^2\rangle=30.3225$~fm$^2$, $\langle r^4\rangle=1169.38$~fm$^4$ for $^{208}$Pb. Meanwhile, the referenced second moments from the RCHB calculations are also shown in the panels. The relative deviations in charge radii read $|\sqrt{12.0778}-\sqrt{12.0755}|/\sqrt{12.0755}\approx0.01\%$ for $^{40}$Ca and $0.02\%$ for $^{208}$Pb. 
\textcolor{black}{Note that the source of the deviation here is the model dependency. From the above numbers, it can be seen that the influence of the model dependency on the extracted charge radii is rather small.}

In practice, the accuracy of the extracted charge radii is limited by the total uncertainty of transition energies from experimental measurements, together with various additional corrections, such as vacuum polarization, relativistic recoil, self-energy, and nuclear polarization corrections~\cite{RevModPhys.54.67,PhysRevA.13.1283,PhysRevA.17.489,PhysRevA.66.034501}. 
\textcolor{black}{Therefore, the determination of uncertainty of the extracted charge radii is in general a challenging task. In the present theory-to-theory benchmarking study, the dependence of the uncertainty in the extracted charge radii $\Delta r_c$ on the uncertainty in the transition energies can be quantitatively investigated. The RMSD $\delta$ in the present calculations can be regarded as the total uncertainty of transition energies,}
and the difference of charge radii $\Delta r_c=r_{c}^\mathrm{2pf}-r_{c}^\mathrm{RCHB}$ can be used to indicate the uncertainty of the extracted charge radii. \textcolor{black}{For this purpose,} the variations of $\delta$ versus $\Delta r_c$ for $^{40}$Ca and $^{208}$Pb are shown in Figs.~\ref{fig:fig2}(c) and \ref{fig:fig2}(f), respectively. The points at $\Delta r_c=\pm0.01$~fm are marked and the corresponding values of $\delta$, i.e., $\delta\approx 0.0017$~keV for $^{40}$Ca and $\delta\approx1$~keV for $^{208}$Pb, are illustrated. In fact, it is quite challenging to make the total uncertainty of transition energies less than 0.0017 keV for $^{40}$Ca.
However, the uncertainty less than $1$~keV for $^{208}$Pb is available~\cite{PhysRevC.37.2821}. As a result, the uncertainty of the extracted charge radii less than $0.01$~fm for $^{208}$Pb can be obtained.
It can be deduced that the charge radii for the heavy nuclei can be extracted from muonic atom spectroscopy more accurately than for the light nuclei.

\begin{figure}
\centering
\includegraphics[width=8.5cm]{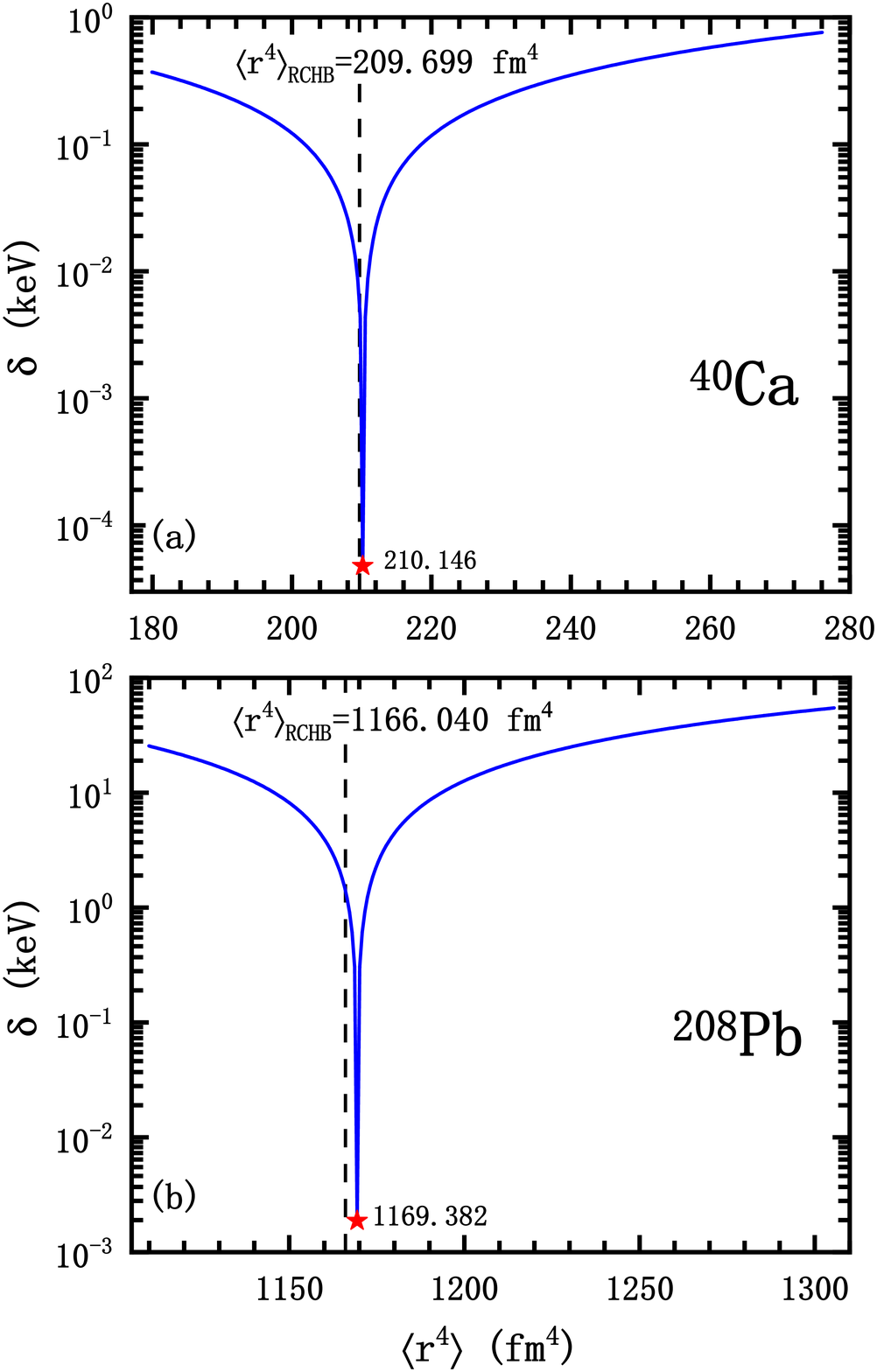}
\caption{The root-mean-square deviation $\delta$ as a function of the fourth moment $R_4$ of 2pf distribution for (a) $^{40}$Ca and (b) $^{208}$Pb. Different from Fig.~\ref{fig:fig2}, an additional constraint is imposed to keep the second moment constant, i.e., $R_2=12.07784~\mathrm{fm^2}$ for $^{40}$Ca and $30.32254~\mathrm{fm^2}$ for $^{208}$Pb. The black dashed lines indicate the data from RCHB calculations.}
\label{fig:fig3}
\end{figure}

In order to study the variation of $\delta$ versus the fourth moment, the influence of the second moment should be avoided. In Fig.~\ref{fig:fig3}, the steep tendency of RMSD $\delta$ with the fourth moment is illustrated, where parameter $t$ is changed from $1$ to $3.5$~fm, and $c$ is determined according to the additional constraint of the second moment. The fourth moment of the RCHB charge distributions is marked and a small deviation from the corresponding value at the point of minimum $\delta$ can be seen. The relative deviations of the fourth moment are only around $0.2\%$ and $0.3\%$ for $^{40}$Ca and $^{208}$Pb, respectively. Therefore, it can be considered that the prediction of the fourth moment from the muonic atom spectrum is reliable.

\begin{figure}[htbp]
\centering
\includegraphics[width=8.5cm]{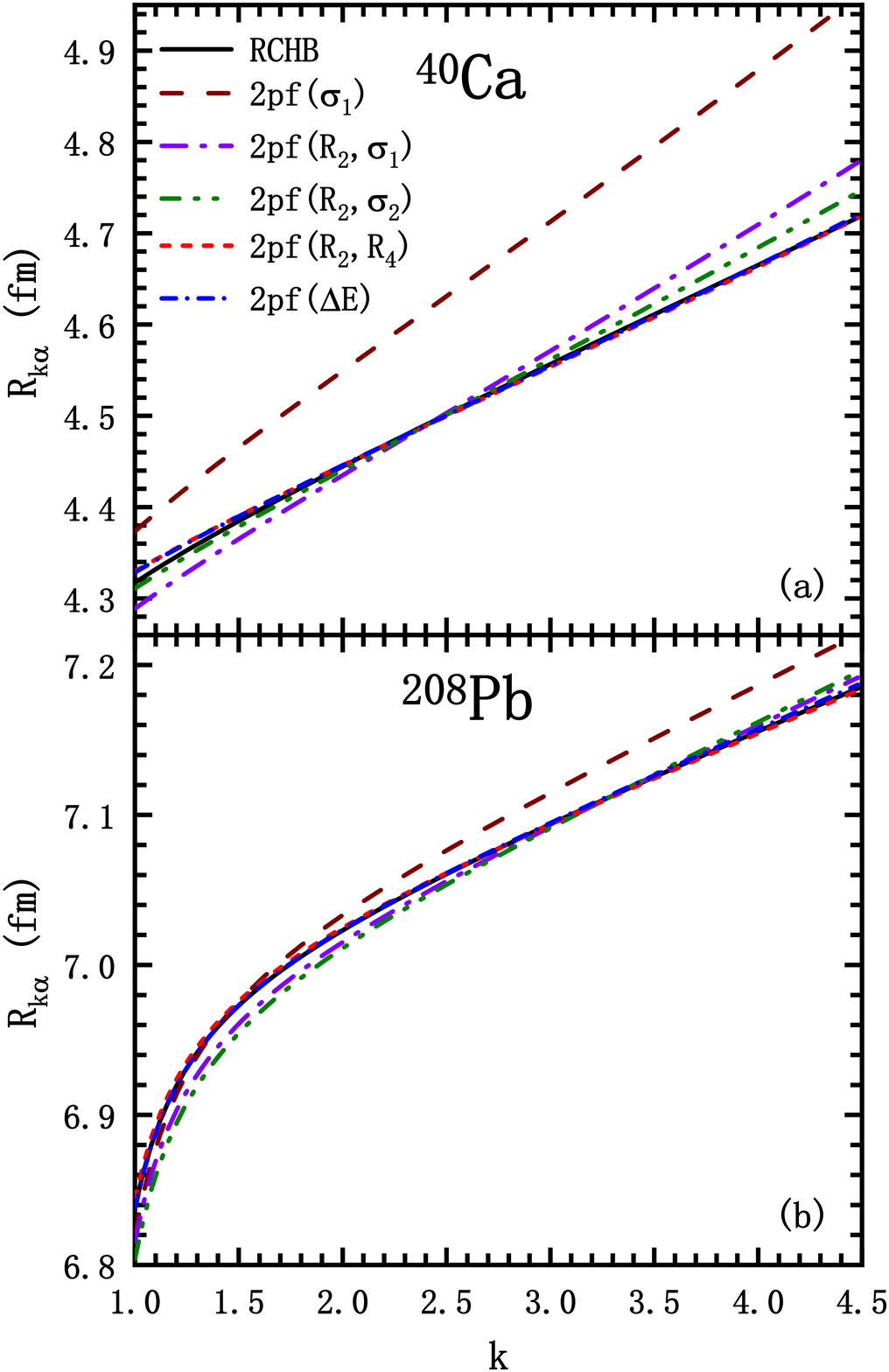}
\caption{The Barrett equivalent radius $R_{k\alpha}$ for (a) $^{40}$Ca and (b) $^{208}$Pb. The values of $\alpha$ in Eq.~(\ref{eq-barrett}) are taken as $0.065$~fm$^{-1}$ for $^{40}$Ca~\cite{PhysRevC.23.533} and $0.1415$~fm$^{-1}$ for $^{208}$Pb~\cite{PhysRevC.37.2821}. }
\label{fig:fig4}
\end{figure}

In addition, the analysis using the Barrett model~\cite{ENGFER1974509} is also performed. The Barrett equivalent radius $R_{k\alpha}$ is defined by
\begin{equation}\label{eq-barrett}
\frac{3}{R_{k\alpha}^3}\int_0^{R_{k\alpha}}r^ke^{-\alpha r}r^2\,\mathrm dr=\langle r^ke^{-\alpha r}\rangle.
\end{equation}
It can be iteratively calculated from the charge density without resolving the Dirac equation.
The critical sensitivity of transition energies in muonic atoms to the Barrett equivalent radius has been illustrated~\cite{BARRETT1970388,FRICKE1995177}. Thus, the Barrett model provides a particular perspective to compare these 2pf distributions with the fitted parameters to the referenced charge density from the RCHB calculations.
The comparisons of the Barrett equivalent radius for four 2pf distributions and the RCHB calculations versus $k$ for $^{40}$Ca and $^{208}$Pb are shown in Fig.~\ref{fig:fig4}. \textcolor{black}{The values of $\alpha$ are in general determined by the fitting of the differences of the potential that muon in state $a$ makes and the potential that muon in state $b$ makes, i.e.,  $f_{ab}(r) = V_a(r) - V_b(r)$~\cite{BARRETT1970388,PhysRevC.37.2821}. The muon potential in the state $a$ is defined by~\cite{ENGFER1974509}
\begin{equation}
  V_a(r)=\int_0^\infty\frac{P^2_a(r')+Q^2_a(r')}{\max(r,r')}\, dr',
\end{equation} 
where $P_a(r)$ and $Q_a(r)$ are the large and small components in Eq.~(\ref{eq-dirac}), respectively. In the present calculations of the Barrett equivalent radius, the values of $\alpha$ are taken as $0.065$~fm$^{-1}$ for $^{40}$Ca~\cite{PhysRevC.23.533} and $0.1415$~fm$^{-1}$ for $^{208}$Pb~\cite{PhysRevC.37.2821}.}
Different from the practical analysis in experimental references, e.g.,~\cite{PhysRevC.37.2821}, the results of benchmark, i.e., ``RCHB'', are displayed in the form of  smooth curves instead of several dots that correspond to transition energies, because they are evaluated from the RCHB charge density instead of the transition energies. It can be seen that the results of 2pf($R_2$,$R_4$) are in excellent agreement with the results of RCHB calculations. In contrast, the results of 2pf$(\sigma_1)$ considerably deviate from the results of RCHB calculations. As for the rest of two 2pf distributions, Fig.~\ref{fig:fig4} shows that the result of 2pf($R_2$,$\sigma_2$) is in a better (worse) agreement with the result of RCHB than  2pf($R_2$,$\sigma_1$) for $^{40}$Ca ($^{208}$Pb), which are consistent with the comparison of transition energies displayed in Table~\ref{tab-2}. 

\textcolor{black}{Furthermore, the results of 2pf($\Delta E$), i.e., the best-fit 2pf distributions for minimizing the RMSD $\delta$ mentioned above, are also compared in the Fig.~\ref{fig:fig4}. It can be seen that these results} are also in excellent agreements with the results of RCHB calculations. It ensures that the parameters fitted to the transition energies give a satisfactory agreement.

\begin{figure}
\centering
\includegraphics[width=8.5cm]{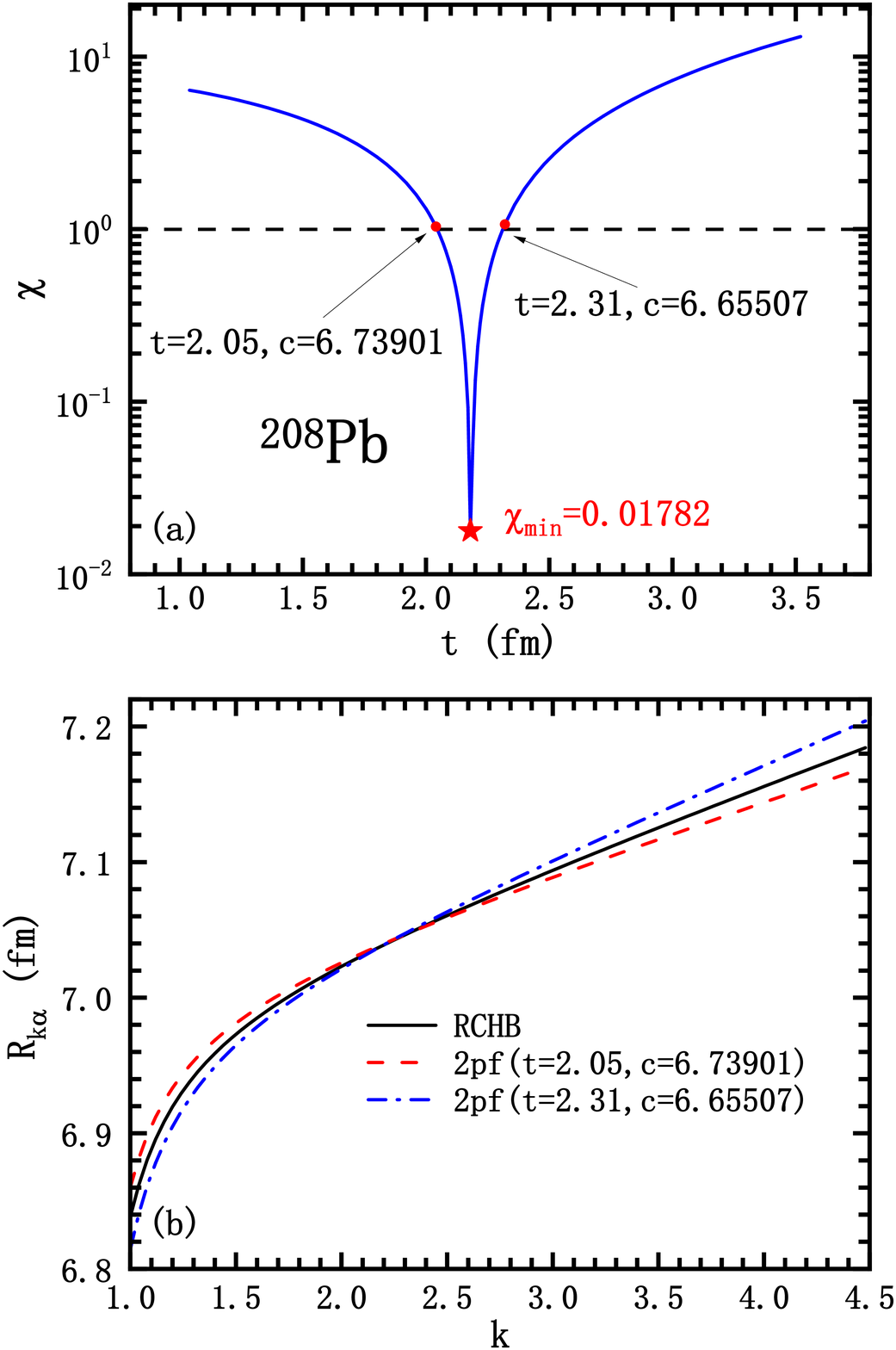}
\caption{
Determination of parameters in 2pf model for $^{208}$Pb. (a) The weight RMSD $\chi$ versus $t$, where $c$ is searched to minimize $\chi$. Two critical points satisfying condition $\chi\le 1$ are indicated. (b) Comparison of the Barrett equivalent radius  between the RCHB calculations and the 2pf distributions with respect to two critical point. 
}
\label{fig:fig5}
\end{figure}

\begin{figure}
\centering
\includegraphics[width=8.5cm]{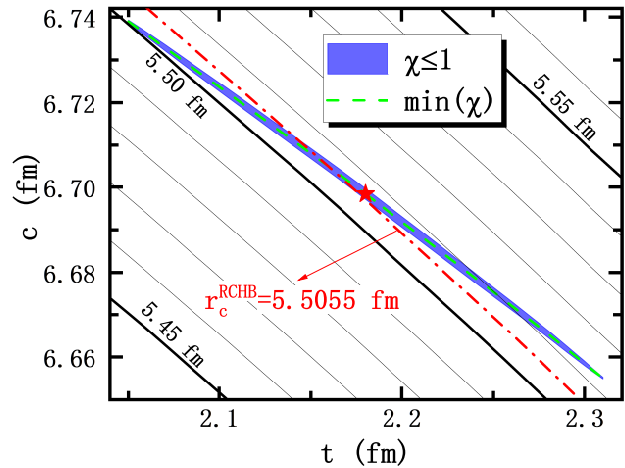}
\caption{
Determination of parameters in 2pf model for $^{208}$Pb.
The elliptic region of 2pf parameters $c$ and $t$ for $\chi\le 1$ is denoted with the blue shadow area, where the red star symbol shows the point with the minimum $\chi$ and the green dashed line shows the determined $c$ values by minimizing $\chi$ for each given $t$.
The contours of the rms charge radii ($0.01$~fm interval) are illustrated with the black solid lines. Additionally, the red dot-dashed line represents the contour of the benchmarking charge radius.}
\label{fig:fig6}
\end{figure}

In order to perform a more practical investigation, we will then focus on the weighted RMSD, which is defined as
\begin{equation}\label{eq-chi-new}
\chi=\left[\frac{1}{N}\sum_i\frac{\left(\Delta E_{i}^\mathrm{2pf}-\Delta E_{i}^\mathrm{RCHB}\right)^2}{\left(\eta_i \Delta E_i^{\rm RCHB}\right)^2}\right]^{1/2},
\end{equation}
where $\eta_i$ represents the relative deviations for $i$-th transition energy.
According to one of the latest  experiments for muonic $^{104}$Pd~\cite{Saito2021}, the relative uncertainties of experimental transition energies are approximately $6.47\times 10^{-5}$, $9.81\times 10^{-5}$, $3.47\times 10^{-4}$, and $1.20\times 10^{-4}$ for transitions $2p_{3/2}$-$1s_{1/2}$,  $2p_{1/2}$-$1s_{1/2}$, $3d_{3/2}$-$2p_{1/2}$, and $3d_{5/2}$-$2p_{3/2}$, respectively. Thus, the corresponding $\eta_i$ in Eq.~(\ref{eq-chi-new}) are taken as $10^{-4}$, $10^{-4}$, $3\times 10^{-4}$, and $3\times10^{-4}$, respectively.

In Fig.~\ref{fig:fig5}, the confidence region that satisfies the condition $\chi\le 1$ for $^{208}$Pb is displayed. 
Parameter $t$ is changed from $1$ to $3.5$~fm with a step $0.01$~fm, and $c$ is determined to minimize $\chi$ for each $t$. Then two critical points could be found and the confidence region $2.05\le t\le 2.31$ is obtained in Fig.~\ref{fig:fig5}(a). This implies that the relative uncertainty of $t$ is around $(2.31 - 2.05)/(2.31 + 2.05) \approx 6\%$, while the relative uncertainty of $c$ reaches around $0.6\%$.

In Fig.~\ref{fig:fig5}(b), the Barrett equivalent radius of the 2pf distributions with parameters corresponding to two critical points are compared with the results of RCHB calculations. 
It is of interest that three curves intersect approximately at a point around $k \approx 2.25$, and the curve of RCHB is sandwiched between two 2pf curves with $k$ in the range from $1$ to $4.5$. 

Furthermore, in Fig.~\ref{fig:fig6}, the elliptic region of 2pf parameters which satisfy the condition $\chi\le 1$ is shown. 
The corresponding charge radii are in the range from $5.5004$ to $5.5129$~fm. Thus, the uncertainty of the extracted charge radii is only around $0.005$~fm, and the relative uncertainty is around $0.11\%$. As a result, it is considered that the extraction of charge radii from the muonic atom spectrum for $^{208}$Pb is reliable.

Meanwhile, there are several interesting observations from the figure. 
First of all, it shows the contour of the rms radius in the $c$-$t$ plane. One can see that the spread of the elliptic region is almost parallel to the contour of the rms radius. Combined with the comparison of relative uncertainty between the parameters $c$ and $t$ and the extracted charge radius as previously discussed, it is confirmed that the transition energy is more sensitive to the rms radius than to the individual parameter $c$ or $t$. See also the discussions in Ref.~\cite{Saito2021}. 
Second, the contour of the referenced rms radius, i.e., $r_c^{\rm RCHB}$, is shown in Fig.~\ref{fig:fig6}.
In addition, in the elliptic region, the point with the minimum $\chi$ and the line with the $c$ determined to minimize $\chi$ for each $t$ is  also shown. They correspond to the red star symbol and the blue curve in Fig.~\ref{fig:fig5}(a), respectively. 
\textcolor{black}{Keeping in mind that, in this figure, the green dashed line (representing the determined $c$ values by minimizing $\chi$ for each given $t$) would be identical with the red dot-dashed line (representing the contour of the benchmarking charge radius), if there is no model dependency on extracting the charge rms radius.
In contrast, it is seen from the figure that the slopes of these two lines are slightly different from each other.
Precisely, the slopes are $-0.32$ and $-0.38$ for the former and latter, respectively.
Such a difference can be used as an indicator of the model dependency.}

\textcolor{black}{Before ending this section, it is also important to mention that one cannot constrain the charge radius of $^{40}$Ca at all, by taking the same relative deviations $\eta_i$.
This is because, to extract the charge radius at the level of $\Delta r_c = \pm 0.01$~fm, the required accuracy of the transition energies is much higher in the case of $^{40}$Ca, as we compared quantitatively in Figs.~\ref{fig:fig2}(c) and \ref{fig:fig2}(f).}

\section{Summary and Perspectives}\label{sec-4}

In this work, the influence of model dependencies on the extraction of charge radii from the muonic atom spectrum and the sensitivities of muonic transition energies to the second and fourth moments of charge distribution are investigated by taking $^{40}$Ca and $^{208}$Pb as examples. The charge densities from the RCHB calculations are used as a benchmark.  
The parameters $c$ and $t$ in the 2pf model are determined in several ways to fit the referenced charge densities, and the corresponding transition energies are compared with the data from RCHB calculations. The sensitivities of transition energies to the second and fourth moments are illustrated.

In detail, first of all, the variations of RMSD $\delta$ of muonic transition energies versus parameters $c$ and $t$ are investigated.
By changing $t$ from $1$ to $3.5$~fm and determining $c$ iteratively to minimize $\delta$ for each $t$, the steep trend of $\delta$ versus the second moment is clearly shown. 
Meanwhile, the variations of $\delta$ versus $\Delta r_c$ are also shown to investigate the dependence of the uncertainty in the extracted charge radii on the uncertainty in the transition energies.
On the one hand, it is found that the total uncertainty of transition energies should be less than $0.0017$~keV for $^{40}$Ca and $1$~keV for $^{208}$Pb to make the uncertainty of extracted charge radii less than $0.01$~fm. 
This illustrates that the charge radii of heavy nuclei could be extracted  more accurately from muonic atom spectroscopy than those of light nuclei.
On the other hand, when the charge radii are extracted by minimizing $\delta$, it is found that the relative deviations of extracted charge radii are around $0.01\%$ for $^{40}$Ca and $0.02\%$ for $^{208}$Pb, compared with the referenced charge radii.
Therefore, it is considered that the model dependency only slightly influences the extraction of charge radii. 
Second, the variation of $\delta$ versus the fourth moment of 2pf distribution is displayed with the constraint of the second moment. The sensitivities of transition energies to the fourth moment can be clearly seen. 
Finally, the analysis with the Barrett model is performed. The Barrett equivalent radii obtained by various 2pf distributions are compared with the referenced results.
It is clearly seen that the 2pf distribution determined by the  transition energies 2pf($\Delta E$) and that with the second- and fourth-moment constraints 2pf($R_2$, $R_4$) reproduce the benchmarking Barrett equivalent radii excellently, whereas those without the fourth-moment constraint 2pf($R_2$, $\sigma_1$) and 2pf($R_2$, $\sigma_2$) provide less satisfactory results, and that only focusing on the density distribution alone 2pf($\sigma_1$) shows the result considerably deviating from the benchmark.

As a step further, the analysis with respect to the weighted RMSD $\chi$ with the relative deviations $\eta_i$ taken from the recent experimental data is performed, by taking $^{208}$Pb as an example. The confidence region for parameters $c$ and $t$, satisfying the condition $\chi\le 1$, is exhibited in the $c$-$t$ plane. The small uncertainty of charge radius, i.e., $0.005$~fm, emphasizes the reliability of the extracted charge radii from the muonic atom spectrum.
Meanwhile, it is considered that the difference of slopes of the two lines---one represents the determined $c$ values by minimizing $\chi$ for each given $t$ and the other represents the contour of the benchmarking charge radius---in the $c$-$t$ plane can be used as an indicator of the model dependency.
However, it is also found that the obtained 2pf distributions cannot reproduce the details of the benchmarking charge densities.
In particular, its surface-diffuseness parameter $t$ cannot be determined accurately with the present experimental uncertainties on the muonic transition energies.

In the present theory-to-theory benchmarking study, we have analyzed the model dependencies and determined the uncertainty of extracted charge radii from muonic atom spectroscopy based on the 2pf distribution. It is also expected to see how much information on the nuclear structure can be extracted from the muonic atom spectroscopy using a model-independent distribution, such as the Fourier-Bessel series expansion~\cite{DREHER1974219}. This question would be of great interest in our future studies.

\begin{acknowledgments}
This work was supported by
the Natural Science Foundation of Jilin Province (No.~20220101017JC),
the National Natural Science Foundation of China (No.~11675063),
the Key Laboratory of Nuclear Data Foundation (JCKY2020201C157),
the JSPS Grant-in-Aid for Early-Career Scientists under Grant No.~18K13549,
the JSPS Grant-in-Aid for Scientific Research (S) under Grant No.~20H05648,
the JSPS Grant-in-Aid for Research Activity Start-up under Grant No.~22K20372,
the JSPS Grant-in-Aid for Transformative Research Areas (A) under Grant No.~23H04526,
the JSPS Grant-in-Aid for Scientific Research (B) under Grant No.~23H01845,
the JSPS Grant-in-Aid for Scientific Research (C) under Grant No.~23K03426,
the RIKEN iTHEMS Program, 
the RIKEN Pioneering Project: Evolution of Matter in the Universe,
and the RIKEN Special Postdoctoral Researchers Program. 
\end{acknowledgments}


%

\end{document}